\documentstyle[12pt,epsfig]{article}
\font\tenrm=cmr10

 1
\font\elevenrm=cmr10 scaled\magstep 1

\renewenvironment{thebibliography}[1]
 { \elevenrm
   \begin{list}{\arabic{enumi}.}
    {\usecounter{enumi}     \setlength{\parsep}{0pt}
     \setlength{\itemsep}{3pt} \settowidth{\labelwidth}{#1.}
     \sloppy
    }}{\end{list}}

\begin{document}
\title{\Large \bf Mathematical aspects of the nuclear glory phenomenon;\\ 
backward focusing and Chebyshev polynomials}

\author{Vladimir B.~Kopeliovich$^{a,b}$\footnote{{\bf e-mail}: kopelio@inr.ru}\\
\small{\em a) Institute for Nuclear Research of RAS, Moscow 117312, Russia} \\
\small{\em b) Moscow Institute of Physics and Technology (MIPT), Dolgoprudny, 
Moscow district 141701, Russia} }
\maketitle
{\rightskip=2pc
 \leftskip=2pc
 \noindent}
{\rightskip=2pc
 \leftskip=2pc

\tenrm\baselineskip=11pt
\begin{abstract}
{The angular dependence of the cumulative particles production off nuclei near the kinematical boundary
for multistep process is defined by characteristic polynomials in angular variables, describing spatial momenta
of the particles in intermediate and final 
states. Physical argumentation,         
exploring the small phase space method, leads to the appearance of equations for these polynomials 
in $cos (\theta/N)$, where $\theta$ is the polar angle of the momentum of final (cumulative) particle, 
the integer $N$ being the multiplicity of the process (the number of interactions).
It is shown explicitly how these equations aappear, and the recurrent relations between polynomials with different $N$ 
are obtained. Factorization properties of characteristic polynomials found previously, are extended,
and their connection with known in mathematics Chebyshev polynomials of 2-d kind is established. 
As a result, differential cross section of the cumulative particle production has characteristic behaviour 
$d\sigma \sim 1/ \sqrt {\pi - \theta}$ near the strictly backward direction ($\theta = \pi$, the backward focusing effect).
Such behaviour takes place for any multiplicity of 
the interaction, beginning with $n=3$, elastic or inelastic (with resonance excitations in intermediate states), and can be called 
the nuclear glory phenomenon, or 'Buddha's light' of cumulative particles.} 
\end{abstract}
 \noindent
\vglue 0.3cm}
\newpage 

\section{Introduction} 
One of directions of studies in high energy nuclear physics are studies of the high energy
particles (nuclei) interactions with nuclei.
Intensive studies of the particles production processes in high energy interactions
of different projectiles with nuclei, in regions forbidden by kinematics for the 
interaction with a single free nucleon, or cumulative particles production, started at 70-th 
mostly in Dubna (JINR), beginning with the paper \cite{baldin1}, and in Moscow (ITEP) \cite{leksin1,leksin2,fran-le-2} 
(some restricted review of data can be found in \cite{koma1,koma2}, and also in \cite{glauber}-\cite{chebkop} ).

The main goal of these studies was to find features or peculiarities of the nuclear structure, which make the nucleus $A$
different from a collection of corresponding number ($A$) of weakly interacting nucleons, as it was established
previously.
The interpretation of these phenomena as being manifestation of internal structure of
nuclei assumes that the secondary interactions, or, more generally, multiple
interactions processes (MIP) do not play a crucial role in such production. 
Generally, the role of secondary interactions in the particles production off nuclei 
is at least two-fold: they decrease the amount
of produced particles in the regions, where it is large (it is, in particular, the screening phenomenon), 
and increase the production probability in regions where it was small;
so, they smash out the whole production picture.

In the case of the large angle particle production the background processes which mask 
the possible manifestations of nontrivial features of nuclear structure, are 
subsequent multiple interactions with nucleons inside the nucleus leading to the
particles emission in the "kinematically forbidden" region (KFR).
The problem of the background is a real problem in most of physical experiments
aimed to find new phenomena. In the case of cumulative particles production studies of
multistep, or cascade processe have been unpopular among physisists \footnote{Moreover, such studies
have been absolutely out of fashion during long period.}, because the main
goal of experiments in this field was to reveal manifestations of new nontrivial
effects in nuclear structure. However, it has been proved  \cite{konkop,kop1,kop2,long,chebkop},
that multistep processes provide not negligible contribution to the cumulative production
cross sections, although other important contributions are not excluded and remain to be the
main purpose of futher studies.

The small phase space method developed previously in \cite{kop1,kop2,long} allows to get analytical expressions for the
probability of the multiple interaction processes near the corresponding kinematical boundaries.
The quadratic form in angular variables (deviations from the optimal kinematics) plays the key role
in this approach. The recurrent relations for the characteristic polynomials in polar angles deviations
have been obtained in \cite{koma2} and are reproduced in present paper. 
The connection of these polynomials with Chebyshev polynomials of 2-d kind,
known in mathematics since middle of 19-th century \cite{cubic,cheb-engl,cheb-r} and used in approximation 
theory, has been established.
It is an example of interest when physics arguments have led to some results in mathematics. 

The aim of present paper is to provide mathematically complete explanation of essential details of the method used in
\cite{kop1,kop2,long}, which were not complete just from mathematical point of view.
In the next section the peculiarities of kinematics of the processes in KFR are recalled,
in section 3 the small phase space method of the MIP contributions calculation to the particles
production cross section in KFR is described according to \cite{kop1,kop2,long,chebkop}. 
In section 4 the characteristic polynomials in polar angle deviations from the optimal kinematics
 which define the angular dependence of the cross section on the emission angle of the final (cumulative) 
particle and lead to the backward focusing effect, similar to the known in optics glory
phenomenon. 
The recurrent relations between polynomials
with different $N$ are obtained, and their connection with Chebyshev polynomials of 2-d kind
(Chebyshev-Korkin-Zolotarev, or CKZ-polynomials) is established.
Some generalizations for the case of inelastic rescatterings are presented as well.
Final section contains discussion of problems and conclusions.  Appendix contains some technical details
which can be useful for careful reading of the paper.

\section{Features of kinematics of the processes in KFR}
When the particle with 4-momentum $k_0$ interacts with the nucleus with the mass $m_t \simeq A m_N$,
and the final particle of interest has the 4-momentum $k_f$ the basic kinematical relation is
$$ (k_0+ p_t - k_f)^2 \geq M_f^2, \eqno (2.1) $$
where $M_f$ is the sum of the final particles masses, except the detected particle of interest.
At large enough incident energy, $\omega_0 = k_0^0 \gg M_f$, we obtain easily
$$ \omega_f - z k_f \leq m_t, \eqno (2.2) $$
which is the basic restriction for such processes. Here $k_f = |\vec k_f|$, $z=cos\,\theta < 0$ for particle produced
in backward hemisphere. The quantity $(\omega_f - z k_f)/m_N$ has been called the cumulative number
(more precize, the integer part of this ratio plus 1).

Let us recall some peculiaeities of the multistep processes kinematics established
first in \cite{kop1,kop2} and described in details in \cite{long}. 

\subsection{Rescatterings}
For light particles (photon, also $\pi$-meson) iteration of the Compton formula
$${1\over \omega_{n} } -{1\over \omega_{n-1}} \simeq {1\over m} \left[1-cos (\theta_n)\right] \eqno (2.3)$$
allows to get the  final energy in the form                     
$$  {1\over \omega_N}- {1\over \omega_0} = {1\over m} \sum_{n=1}^N  \left[1-cos (\theta_n)\right] \eqno (2.4)  $$

The maximal energy of final particle is reached for the coplanar process when
all scattering processes take place in the same plane and each angle equals to
$\theta_k=\theta/N$.
As a result we obtain
$$  {1\over \omega_N^{max}}- {1\over \omega_0} = {1\over m} N \left[1-cos (\theta/N)\right] \eqno (2.5) $$
Already at $N>2$ and for $\theta \leq \pi$ the expansion can be made
$$ 1-cos (\theta/N) \simeq \theta^2/2N^2 \eqno (2.6)$$
and for large enough $\omega_0$ we obtain
$$\omega_N^{max} \simeq N {2m\over \theta^2}. \eqno (2.7)$$
This means that the kinematically forbidden for interaction with one nucleon region
is partly filled up due to elastic rescatterings.
Remarkably, that this rather simple property of rescattering processes has not been
even mentioned in pioneer papers \cite{baldin1} - \cite{leksin2}
\footnote{This property was well known, however, to V.M.Lobashev, who observed experimentally
that the energy of the photon after 2-fold interaction can be substantially
greater than the energy of the photon emitted at the same angle in
1-fold interaction.}.

In the case of the nucleon-nucleon scattering (scattering of particles with equal 
nonzero masses in general case) it is convenient to introduce the factor
$$ \zeta = {p\over E+m},     \eqno (2.8)  $$
where $p$ and $E$ are spatial momentum and total energy of the particle
with the mass $m$. When scattering takes place on the particle which is at rest in
the laboratory frame, the $\zeta$ factor of scattered particle is multiplied
by $cos\,\theta$, where $\theta$ is the scattering angle in the laboratory frame.
So, after $n$ rescatterings we obtain the $\zeta$ factor
$$\zeta_N = \zeta_0 cos\,\theta_1 cos\,\theta_2 ... cos\,\theta_N. \eqno (2.9)$$
As in the massless case, the maximal value of final $\zeta_N$ is obtained when
all scattering angles are equal
$$ \theta_1 = \theta_2=...=\theta_N =\theta/N, \eqno (2.10)$$
and the process proceeds in one plane.  So, we have
$$\zeta_N^{max} = \zeta_0 \left[cos (\theta/N)\right]^N.\eqno (2.11) $$

The final momentum is from here
$$ p^{max} = 2m {\zeta ^{max}\over 1-\left(\zeta^{max}\right)^2}. \eqno (2.12)$$
Again, at large enough $N$ and large incident energy ($\zeta_0 \to 1 $) the expansion can be made
at $k\gg m$, and we obtain
$$p_N^{max} \simeq N {2m\over \theta^2} \eqno (2.13)$$
which coincides with previous result for the rescattering of light particles. 
However, the preasymptotic corrections to this result are greater than in former case of the light particle.

The normal Fermi motion of nucleons inside the nucleus makes these boundaries wider \cite{long}:
$$p_N^{max} \simeq N {2m\over \theta^2} \left[1 + {p_F^{max}\over 2m}\left(\theta + {1\over \theta} \right) \right], \eqno (2.14)$$
at $\theta \sim \pi$, where it is supposed that the final angle $\theta$ is large. For numerical estimates we
took the step function for the distribution in the fermi momenta of nucleons inside
of nuclei, with $p_F^{max}/m\simeq 0.27$ \cite{long} and references there.

There is characteristic decrease (down-fall) of the cumulative particle production cross
section due to simple rescatterings, near the strictly backward direction. 
However, inelastic processes with excitations of intermediate
particles, i.e. with intermediate resonances, are able to fill up the region at $\theta \sim \pi$.

\subsection{Resonance excitations in intermediate states}
The elastic rescatterings themselves are only the "top of the iceberg".
Excitations of the rescattered particles, i.e. production of resonances in intermediate
states which go over again into detected particles in subsequent interactions,
provide the dominant contribution to the production cross section.
Simplest examples of such processes may be $NN \to NN^* \to NN$, $\pi N\to \rho N \to \pi N$,
etc.

To produce the final particle near the absolute boundary one needs to have the masses of intermediate 
resonances (or some other object)  of the order of incident energy, $s\sim k_0 m_A$.
As a first example let us consider the interaction with the deuteron.

The mass of the object in intermediate state equals
$$M^2_1 =(k_0-k_2)d/2 +(k_0^2+k_2^2)/2 +p_2^2-d^2/4,  \eqno (2.14) $$
where $k_0$ and $k_2$ are the 4-momenta of the incident and final particles,
$p_2$ is the 4-momentum each of final nucleons, $d$ - the deuteron 4-momentum.
The motion of nucleons inside the deuteron leads to considerable decrease of the
mass $M_1$ necessary to produce the particle at the absolute kinematical boundary.

It is not difficult to find the necessary masses of intermediate states in the extreme case
when the emitted particle is just on the absolute boundary for the interaction with 
arbitrary nucleus of atomic number $A$ \cite{kop2,long}. 
Calculation of contributions of such processes
to the production cross section of cumulative particle of interest is not possible due to lack of
information about elementary processes amplitudes of resonances excitation - deexcitation at necessary energies and
momentum transfers.

What is very important: at arbitrary high incident energy the kinematics of all
subsequent processes is defined by the momentum and the angle of the outgoing particle.
The number of different processes (with different resonances excited in intermediate states) is very large
and growes exponentially with increasing $N$ - the number of interactions (like $(N_R +1)^{(N-1)}$, where $N_R$ is
the number of resonances. Therefore, for any final 
(cumulative paarticle) momentum $k$ at any emission angle $\theta $ there are processes, which kinematical boundary
is just near this value of $k$, and our consideration can be applied. \section{The small phase space method for the MIP probability calculations}

This method, most adequate for anlytical and semi-analytical calculations of
the MIP probabilities, has been proposed in \cite{kop2} and developed later in 
\cite{long}. It is based on the fact that, according to established in \cite{kop1,kop2}
and presented in previous section kinematical relations, there is a preferable plane of the 
whole MIP leading to the production of energetic particle at large angle $\theta$, 
but not strictly backwards. Also, the angles of subsequent rescatterings are close
to $\theta / N$. Such kinematics has been called optimal, or basic kinematics.
The deviations of real angles from the optimal values are small, they are defined mostly
by the difference $k_N^{max} - k $, where $k_N^{max}(\theta)$ is the maximal possible momentum reachable
for definite MIP, and $k$ is the final momentum of the detected particle.
$k_N^{max}(\theta)$ should be calculated taking into account normal Fermi
motion of nucleons inside the nucleus, and also resonances excitation ---
deexcitation in the intermediate state. Some high power of the difference  $(k_N^{max} - k)/k_N^{max} $
enters the resulting probability.

Within the quasiclassical treatment adequate for our case, the probability product approximation is
valid, and the following starting expression for the inclusive cross section of the particle
production at large angles takes place (see, e.g., Eq. (4.11) of \cite{long}):

$${d^3k\over \omega} f_N = \pi R_A^2 G_N(R_A,\theta) \int {f_1(\vec k_1) d^3k_1\over \sigma^{leav}_1 \omega_1} \prod_{l=2}^N \frac{M_k^2(s_k,t_k) 
\delta(m +\omega_{l-1}-\omega_l-\omega_{l-1})}{(8\pi)^2 \sigma^{leav}_l m k_{l-1} \omega_l\omega_{l-1}} d^3k_l \eqno(3.1) $$
Here $\sigma_l^{leav}$ is the cross section defining the removal (or leaving) of the rescattered object at the corresponding section
of the trajectory. It includes all inelastic cross section, the part of elastic cross section and the part of the resonance
production cross sections, and can be considerably smaller than the total interaction cross section of the $l$-th intermediate
particle with nucleon. $G_N(R_A,\theta)$ is the geometrical factor defining the probability of the $N$-fold interaction with
definite trajectory of the interacting particles (resonances) inside the nucleus. This trajectory is defined mostly
by the final values of $\vec k, \,\theta $, according to the kinematical relations of previous section. $f_1 = \omega_1 d^3\sigma_1/d^3k_1$,
$\omega'_N=\omega$ --- the energy of the observed particle.

After some evaluation and introducing the differential cross sections of binary reactions $d\sigma_l/dt_l(s_l,t_l) $ instead of 
the matrix elements of binary reactions $M_l^2(s_l,t_l)$, we came to the formula for the production cross section 
due to the $N$-fold MIP \cite{kop2,long}
$$f_N(\vec p_0,\vec k)= \pi R_A^2 G_N(R_A,\theta) \int \frac{f_1(\vec p_0,\vec k_1) (k_1^0)^3 x_1^2dx_1 d\Omega_1}{\sigma_1^{leav}\omega_1}
\prod_{l=2}^N\left({d\sigma_l(s_l,t_l)\over dt}\right) \frac{(s_l-m^2-\mu_l^2)^2-4m^2\mu_l^2}{4\pi m\sigma_l^{leav} k_{l-1}} $$
$$\times \prod_{l=2}^{N-1}\frac{k_l^2 d\Omega_l}{k_l(m+\omega_{l-1}-z_l\omega_lk_{l-1})\;}{1\over \omega_N'}\delta(m+\omega_{N-1}-\omega_N-\omega_N').
\eqno(3.2) $$ 

\subsection{Rescattering of the light particle ($\pi$-meson)}
Further details depend on the particular process. For the case of the light particle rescattering, $\pi$-meson for example, $\mu_l^2/m^2\ll 1$,
we have 
$${1\over \omega_N'}\delta(m+\omega_{N-1}-\omega_N-\omega_N') = {1\over k k_{N-1}}
\delta\left[ {m\over k} - \sum_{l=2}^N (1-z_l) - {1\over x_1}\left({m\over p_0}+1-z_l\right)\right] \eqno(3.3) $$
To obtain this relation, one should use the equality (energy-momentum conservation in the last interaction act)
$$ \omega_N' = \sqrt{m^2 + (\vec k_{N-1}^2 - \vec k)^2 } $$
and the known rules for manipulations with the $\delta$-function, see Appendix.
When the final angle $\theta$ is considerably
different from $\pi$, there is a preferable plane near which the whole
multiple interaction process takes place, and only processes near this plane
contribute to the final output. At the angle $\theta =\pi$, strictly backwards,
there is azimuthal symmtry, and the processes from the whole interval of azimuthal 
angle $\phi$,  defining the plane of the process, $0< \phi < 2\pi$, provide contribution to the final output (azimuthal focusing, see next section).
A necessary step is to introduce azimuthal  deviations from this optimal
kinematics, $\varphi_k$, $k=1,\,...,N-1$.  $\varphi_N=0$ by definition of the plane of the process, $(\vec p_0, \vec k)$.
Polar deviations from the basic values, $\theta/N$, are denoted as $\vartheta_k$, obviously, 
 $\sum_{k=1}^N\vartheta_k = 0$. The direction of the momentum $\vec k_l$ after $l$-th 
interaction, $\vec n_l$,  is defined by the azimuthal angle $\varphi_l$ and the polar angle 
$\theta_l = (l\theta /N) +\vartheta_1+...+\vartheta_l$.

Then we obtained \cite{kop2,long} making the expansion in $\varphi_l$,  $\vartheta_l$ and including quadratic terms
in these variables:
$$z_k= (\vec n_k \vec n_{k-1}) \simeq cos (\theta/N) (1-\vartheta_k^2/2) -sin (\theta/N) \vartheta_k 
+ sin (k\theta/N) sin [(k-1)\theta/N] (\varphi_k -\varphi_{k-1})^2/2. \eqno (3.4)$$
In the case of the rescattering of light particles the sum enters the phase space of the process
$$ \sum_{k=1}^N (1- cos\vartheta_k) =  N[1-cos(\theta/N)] + cos(\theta/N)
\sum_{k=1}^N\bigg[-\varphi_k^2\, sin^2(k\theta/N) + $$
$$+{1\over cos(\theta/N)}sin(k\theta/N)sin((k-1)\theta/N)\bigg] -{cos(\theta/N)\over 2} \sum_{k=1}^N \vartheta_k^2 \eqno (3.5)$$ 
To derive this equality we used that $\varphi_N=\varphi_0=0$ --- by definition of the plane of the MIP,
and the mentioned relation $\sum_{k=1}^N\vartheta_k = 0$.
We used also the identity,  valid for $\varphi_N=\varphi_0 =0$:
$${1\over 2}\sum_{k=1}^N \left(\varphi_k^2+\varphi_{k-1}^2 \right) sin(k\theta/N)sin[(k-1)\theta/N] = 
cos(\theta/N)\sum_{k=1}^N\varphi_k^2 sin^2(k\theta/N). \eqno (3.5a)$$
It is possible to present it in the canonical form and to perform integration easily, see
Appendix and Eq. $(4.23)$ of \cite{long}.
As a result, we have the integral over angular variables of the following form:
$$ I_N(\varphi, \vartheta) = \int \delta\biggl[\Delta^{ext} - z_N\bigg(\sum_{k=1}^{k=N} \varphi_k^2 -\varphi_k\varphi_{k-1}/z_N
+\vartheta_k^2/2\biggr)\biggr] \prod_{l=1}^{N-1} d\varphi_ld\vartheta_l = $$
$$= \frac{\left(\Delta^{ext}\right)^{N-2} (\sqrt 2 \pi)^{N-1}}{J_N(z_N) \sqrt N (N-2)! z_N^{N-1}}, \eqno (3.6) $$
Since the element of a solid angle $d\Omega_l= sin(\theta\,l/N)d\vartheta_l d\varphi_l $, we made here substitution
$sin(\theta\,l/N)\,d\varphi_l \to d\varphi_l$.
$z_N= cos(\theta/N)$, $\Delta^{ext}\simeq m/k - m/p_0 -N(1 - z_N) +(1-x_1)m/p_0 $.

$\Delta_N^{ext}$ defines the distance of the momentum (energy) of the emitted particle $\vec k, \, \omega$ from the kinematical boundary for 
the whole $N$-fold MIP.
$$ J_N^2(z) = Det\, ||a_N||, \eqno (3.7)$$
where the matrix $||a||$ defines the quadratic form $Q_N(z)$ which enters the argument of the $\delta$-function in Eq. $(3.6)$:
$$ Q_N(z,\varphi_k) = a_{kl} \varphi_k \varphi_l =\sum_{k=1}^{k=N} \varphi_k^2 -{\varphi_k\varphi_{k-1}\over z}. \eqno (3.8)$$
 For example:
$$Q_2 =\varphi_1^2, \quad  Q_3 = \varphi_1^2 +\varphi_2^2 -{\varphi_1\varphi_2 \over cos(\theta/3)}; \quad 
Q_4 = \varphi_1^2 +\varphi_2^2 +\varphi_3^2-{\varphi_1\varphi_2 \over cos(\theta/4)} -  {\varphi_2\varphi_3 \over cos(\theta/4)}; ...$$
Properties of these quadratic forms are considered in the next section.

The phase space of the process $(3.3)$ which depends strongly on $\Delta_N^{ext}$ after integration over angular 
variables takes the form
$$\Phi_N^{pions} = {1\over \omega_N'}\delta(m+\omega_{N-1}-\omega_N-\omega_N')\prod_{l=1}^Nd\Omega_l =
\frac{(\sqrt 2\pi)^{N-1}(\Delta^{ext})^{N-2}}{kk_{N-1}(N-2)!\sqrt N J_N(z_N)z_N^{N-1}} \eqno(3.15) $$

\subsection{Rescattering of the nucleon}
For the case of nucleons rescattering there are some important differences from the light particle case, but
the quadratic form which enters the phase space of the process is essentially the same:
$$\Phi_N^{nucleons} = {1\over \omega_N'}\delta(m+\omega_{N-1}-\omega_N-\omega_N')\prod_{l=1}^Nd\Omega_l =
\int \delta\left[\Delta^{ext}_{nucl} -z_N^N Q_N(\varphi_k) -{z_N^{N-2}\over 2}\sum_{l=1}^N\vartheta_l^2\right] =$$
$$=\frac{(\sqrt 2\pi)^N(\Delta^{ext})^{N-2}}{kk_{N-1}(N-2)!\sqrt N J_N(z_N)z_N^{N-1}} \eqno(3.16) $$

The normal Fermi motion of target nucleons inside of the nucleus increases the phase space considerably \cite{kop2, long}:
$$\Delta^{ext} = \Delta^{ext}|_{p_F=0} + \vec p^F_l\vec r_l/2m, $$
where $\vec r_l =2m(\vec k_l-\vec k_{l-1})/k_lk_{l-1} $. A reasonable approximation is to take vectors $\vec r_l$ according to
the optimal kinematics for the whole process, and the Fermi momenta distribution of nucleons inside of the nucleus
in the form of the step function.

For the case of the nucleons rescattering some difference from the case of the light particle rescattering takes place,
but the axial focusing effect persists.
$$ \Phi_N = {1+ \zeta_1^2 \over k(m+\omega_{N-1})\zeta_N(1-\zeta_1^2)} \left( {\sqrt 2 \pi\over \zeta_{N-1}}\right)^{N-1}
\frac {[\zeta_N - k/(\omega +m)]^{N-1}}{I_N(z) \sqrt N (3N-1)!} \eqno (3.17) $$

We obtained for this case, taking into account the normal Fermi motion of nucleons in the nucleus 
( \cite{kop2} and Eq. (4.25) of \cite{long}):
$$ \Phi_N^{nucleons} = \int {1\over \omega_N'} \delta(m+\omega_{N-1}-\omega_N-\omega_N')\prod_{l=1}^{N}d\Omega_l \rho(\vec p_{Fl})d^3p_{Fl}=$$ 
$$=\frac{z_N^{N^2+N-1}(1+\zeta_0^2z_N^2)(\sqrt 2 \pi)^{N-1}(3/2b^2)^N(\Delta_N^{nucl})^{3N-1}}{k(m+\omega_{N-1})J_N(z_N) \zeta_0^{N-1}
(1-\zeta_0^2 z_N^2)\sqrt N (3N-1)!} \prod_{i=1}^N\frac{1- (\Delta_N^{nucl})z_N^i/(3Nbr_i^{nucl})}{(r_i^{nucl})^2} \eqno (3.18) $$
with $b=p_F^{max}/2m,\; r_i^{nucl}= 2m |\vec k_i -\vec K_{i-1}|/k_i(m+\omega_{i-1})$.
At high enough incident energy substitution $\zeta_0 \to 1$ can be done.

\section{Quadratic form in angular deviations, characteristic polynomials and their properties}
\subsection{Relations between characteristic polynomials}
The obvious recurrent relation takes place for the quadratic form in azimuthal deviations:
$$Q_{N+1}(z,\varphi_k,\varphi_l)  = Q_N(z,\varphi_k,\varphi_l) +\varphi_N^2 -\varphi_N\,\varphi_{N-l}/z, \eqno (4.1)$$
where $z=cos[\theta /(N+1)]$, has the same value in both sides of this equation, $\varphi_{N+1} =0$ by definition 
of the plane of the process.  Relation $(4.1)$ plays a key role to establish the connection of characteristic polynomials
$(4.3),\,(4.5)$ with Chebyshev polynomials of 2-d kind, noted at first in \cite{chebkop}.

Let $t$ be the transformation (matrix) which brings our quadratic form to the canonical form:
$$ \tilde t\, a\, t = {\cal I}, \eqno(4.2)$$
where ${\cal I}$ is the unit matrix $n\times n$, and $\tilde t_{kl} = t_{lk}$.
Then the equality takes place for the Jacobian of this transformation
$$ (det\,||t||)^{-2} = J_a^2(z) = det\,||a||, \qquad (det\,||t||)^{-1} = J_a(z) = \sqrt{det\,||a||}. \eqno(4.3)$$

The matrices $||a_{ij}||^{N}$ can be presented in the following symmetric form:
$$||a_{ij}||^{N=3}=\left| \begin{array}{cc} 1&-{1\over 2 cos(\theta/3)} \\-{1\over 2 cos(\theta/3)}& 1  \end{array}\right|,\qquad
Det ||a_{ij}||^{N=3} = 1 - {1\over 4 cos^2 \theta/3}, \eqno (4.4)$$  \\

$$||a_{ij}||^{N=4}=\left| \begin{array}{ccc} 1&-{ 1\over 2 cos(\theta/4)} &0 
\\ -{ 1\over 2 cos(\theta/4)} & 1 & -{ 1\over 2 cos(\theta/4)}
 \\0& -{ 1\over 2 cos(\theta/4)}&1  \end{array}\right|, \eqno (4.5)$$  \\

$$ Det ||a_{ij}||^{N=4} =1-{1\over 4 cos^2\theta/4} + {1\over 2\,cos\theta/4}\left(-{1\over 2\,cos\theta/4}\right)= 1 - {1\over 2\,cos^2\theta/4}. \eqno(4.6)$$

For arbitrary $N$ the following general expression for the matrix $||a_{ij}||^{N}$ can be written:
$$||a_{ij}||^{N}=\left| \begin{array}{ccccccc} 1  &-{ 1\over 2 cos(\theta/N)} & 0 & ... & 0 & 0 & 0
                            \\ -{ 1\over 2 cos(\theta/N)} & 1 & -{ 1\over 2 cos(\theta/N)} & ... & 0 & 0 & 0
\\ ... & ... & ... & ...  & ... & ... & ...
\\  0 & 0    & 0   & ...  &  1  & -{ 1\over 2 cos(\theta/N)}  &  0
\\0 & 0& 0 & ... & -{ 1\over 2 cos(\theta/N)} & 1  &-{ 1\over 2 cos(\theta/N)} 
 \\0&0 & 0 & ... & 0 &  -{ 1\over 2 cos(\theta/N)}&1  \end{array}\right|  \eqno (4.7 )$$\\

According to well known rules for the calculations of the determinants of matrices, the relation can be easily
established

$$ Det ||a_{ij}||^{N} = J_N^2 = J_{N-1}^2 - {1\over 4 cos^2(\theta/N)} J^2_{N-2} \eqno (4.8) $$

Using this recurrent relation, it is possible to present the quadratic form  in $\phi_k, \phi_l$, which enters the $\delta$ - function as
the sum of squares of certain combinations of deviations $\varphi_k$, which will be convenient in future applications:
$$ Q_N(\varphi_k,\varphi_l) =J_2^2\left(\varphi_1-{\varphi_2\over 2zJ_2^2}\right)^2 +{J_3^2\over J^2_2}
\left(\varphi_2-{J_2^2\varphi_3\over 2zJ_3^2}\right)^2 +...$$
$$...  + {J_{N-1}^2\over J_{N-2}^2}\left(\varphi_{N-2}-{J_{N-2}^2\varphi_{N-1}\over 2z J_{N-1}^2}\right)^2 +
{J_N^2\over J_{N-1}^2} \varphi_{N-1}^2 \eqno (4.9)$$
with $J_2^2=1$.

The following general formula for $J_N^2(z_N)$ has been obtained in \cite{long}, Eq. (4.23) \footnote{In the paper \cite{kop2}, Eq (15) this
formula has been presented for $N$ up to $N=5$.}:
$$Det ||a_{kl}|| =J_N^2(z_N) = J_N^2(\theta/N)= 1 + \sum_{m=1}^{m < N/2}\left(-{1\over 4z_N^2}\right)^m 
{\prod_{k=1}^{m}(N-m-k)\over m!} = $$
$$= 1 + \sum_{m=1}^{m < N/2}\left(-{1\over 4z_N^2}\right)^m C^m_{N-m-1}, \eqno (4.10)$$
$z_n=cos (\theta/n)$, $Det\, ||a_{kl}||$ is the determinant of the matrix $||a||$, $C_n^m$ is the number of combinations.

As it became clear to us recently, the polynomials $(4.7)$ coincide, up to some factor depending on $z=cos (\theta/N)$
with Chebyshev polynomials of 2-d kind, discovered in the middle of 19-th century \cite{cubic, cheb-engl, cheb-r}.
The connection of characteristic polynomials $J^2_N\left[cos(\theta/N)\right]$ with Chebyshev polynomials of 2-d kind
(CKZ-polynomials) will be described in next section.

The angle  $\theta_1$ - the polar angle of the momentum of the particle in intermediate state - is different 
from $\theta/N$ if the resonance excitation takes place. It can take place in the first interaction act, and in
any of subsequent interqctions. But the optimal kinematics is changed in any of these processes.
We can write then $\theta_1 = \theta/N + \delta_1$, assuming that $\delta_1\ll \theta/N$.

\subsection{Algebra of characteristic polynomials}

The condition $J_N(\pi/N) =0$ leads to the equation for $z_N$ which solution
(one of all possible roots) provides the value of $cos(\pi/N)$ in terms of radicals.
The following expressions for these jacobians take place \cite{kop2}:
$$J_2^2(z)=1; \quad J_3^2(z) = 1-{1\over 4z^2}; \quad J_4^2(z)= 1-{1\over 2z^2}, \quad J_5^2=1-{3\over 4z^2}+{1\over 16 z^4}\eqno (4.11) $$
$J_3(\pi/3)=J_3(z=1/2)=0$, 
$J_4(\pi/4)=I_4(z=1/\sqrt 2)=0$, $cos^2(\pi/5) = (3+\sqrt 5)/8$.

The case $N=2$ is a special one, because $J_2(z)=1$ - is a constant. But in this case the 2-fold
process at $\theta =\pi$ (strictly backwards) has no advantage in comparison with the direct one,
if we consider the elastic rescatterings.

For $N=5$

$$ J_5^2=1 -{3\over 4z^2} + {1\over 16 z^4}, \eqno (4.12) $$

Several other examples were given later in \cite{long}, and recently in \cite{koma1,koma2}.

At $N=6$
$$J_6^2=1-{1\over z^2}+{3\over 16 z^4} = J_3^2 \left(1-{3\over 4z^2}\right) . \eqno (4.13)$$

For $N=7$
$$J_7^2=1-{5\over 4z^2}+{3\over 8z^4} -{1\over 64z^6}. \eqno (4.14)$$
$J_7(\pi/7)=0$.
One of the solutions of equation $J_7^2(z)=0$ is $z=cos(\pi/7)$, which can be verified using the method by Ferro and Tartaglia (published
by Cardano, see e.g. \cite{cubic} ). Other two solutions can be found easily, when we devide the
polynomial $J_7^2(z)$ by $z-cos(\pi/7)$.

$$J_8^2=1-{3\over 2z^2}+{5\over 8z^4} -{1\over 16z^6}=J_4^2 \left(1-{1\over z^2}+{1\over 8z^4}\right), \eqno (4.15)$$

  $J_8(\pi/8)=0$.
For arbitrary $N$,  $J_N^2$ is a polynomial in $1/4z^2$ of the power $|(N-1)/2|$ (integer part of $(N-1)/2$.
Since the solutions of the equations $J_N^2(z)=0$ are not known in general form when the power of the polynomial is
greater than $5$, the knowledge of at least one solution, $z=cos(\pi/N)$ can be helpful.

The relation can be obtained from here
$$J_N^2(z) = J_{N-k}^2(z)J_{k+1}^2(z)-{1\over 4z^2}J_{N-k-1}^2(z)J_k^2(z) \eqno(4.16)$$
which, at $N=2m,\;k=m$ ($m$ is the integer),  leads to remarkable relation
$$ J_{2m}^2(z)=  J_m^2(z)\left(J_{m+1}^2(z) -\,{1\over 4z^2} J_{m-1}^2(z) \right) =
J_m^2(z)\left(J_{m}^2(z) -\,{1\over 2z^2} J_{m-1}^2(z) \right). \eqno(4.17) $$

TThis relation  can be rewritten in another form, convenient for further investigations:
$$J_{m+n}^2(z) = J_{m}^2(z)J_{n+1}^2(z)-{1\over 4z^2}J_{m-1}^2(z)J_n^2(z) = J_{m+1}^2(z)J_{n}^2(z)-{1\over 4z^2}J_{m}^2(z)J_{n-1}^2(z)\eqno(4.18)$$
This relation  can be verified easily for $J_4^2,\;J_6^2$ and $J_8^2$,  see section 4.
It follows from here that at $N=2m$ not only $J_N(\pi/N) =0$, but also
$J_N(2\pi/N)=0$ which has quite simple explanation ($m\geq 3$; the case of $m=2$ is an exception).

For $N=3m$ we obtain
$$ J_{3m}^2(z)= J^2_m\Biggl\{\left(J^2_{m}\right)^2\left(1-{1\over 4z^2}\right) -{3\over 4z^2}J^2_{m}J^2_{m-1} + 
+{3\over 16z^4}\left(J^2_{m-1}\right)^2 \Biggr\} \eqno (4.19) $$

or

$$ J_{3m}^2(z)= J^2_m\Biggl\{\left(J^2_{m+1}\right)^2 -{1\over 4z^2}\left[J^2_{m-1}J^2_{m+1} + \left(J^2_{m}\right)^2\right]
+{1\over 16z^4}\left(J^2_{m-1}\right)^2 \Biggr\}. \eqno (4.20) $$

For arbitrary odd values of $N$ another useful factorization property takes place:
$$ J_{2m+1}^2(z)= \left(J_{m+1}^2(z)\right)^2-{1\over 4z^2}\left(J_{m}^2(z)\right)^2 =\left(J^2_{m+1}(z) -{1\over 2z} J_{m}^2(z) \right)  
\left(J^2_{m+1}(z) +{1\over 2z} J_m^2(z) \right), \eqno(4.21) $$
which can be easily verified for $J_7^2$ and $J_5^2$ given in section 4.

Evidently, it can be shown by induction, that if $N=p m$ - the product of two integers $p$ and $m$, then
$$J^2_{pm} \sim J^2_m,$$
Indeed, we have from relation $(4.19)$ at $N=pm, \,k=m-1$:
$$J^2_{pm} = J^2_m J^2_{pm-m+1} -{1\over 4z^2} J^2_{m-1} J^2_{(p-1)m}.    \eqno(4.22)) $$
It follows from this equality, that if $J^2_{(p-1)m} \sim J^2_m$, $p-1 >1$, then $J^2_{pm} \sim J^2_m$ as well.
Obviously, at the same time
$$J^2_{pm} \sim J^2_p.$$

This means, that $J^2_{pm} =0 $ not only for $z= cos (\pi/ pm)$, but also for $z=cos (\pi/p)$ and for $z = cos (\pi/m)$.
This factorization property allows to simplify the calculation of the roots of these polynomials.

The connection of polynomials $J_N^2(z)$ with Chebyshev polynomials of 2-d kind, or CKZ-polynomials (Chebyshev-Korkin-Zolotarev, see further), 
explains naturally  their factorization properties.

Some further examples are of interest.

$$J_{12}^2 = J_3^2 J_4^2 \left(1- {3\over 4z^2}\right) \left(1 -{1\over z^2} +{1\over 16 z^4}\right).  \eqno(4.23) $$
$cos (\pi/12)$ can be found as a root of the equation
$ \left(1 -{1/ z^2} +{1/ 16 z^4}\right) =0 $
which can be solved easily, $cos(\pi/12) =\sqrt{2+\sqrt 3 }/2$.

Another simple example:

$$J_{15}^2 =  J^2_5\Biggl\{(J^2_6)^2 -{1\over 4z^2}\Biggl[ J^2_4J^2_6  +J_9^2  \Biggr] \Biggr\}. \eqno(4.24)    $$

Now we can use the above relations $J^2_6 = J^2_3 (1-3/(4z^2)$ and $J^2_9 = J^2_3(1-3/(2z^2)+9/(16 z^4) -1/(64z^6) $
to obtain
$$J^2_{15} = J^2_3J^2_5 \left(1 -{9\over 4z^2} + {13 \over 8 z ^4} -{3\over 8z^6} +{1\over 256 z^8}\right) \eqno(4.25) $$

Simple explanation of the properties of polynomials $J^2_N$ is presented in the next section.

\section{Connection with Chebyshev polynomials of 2-d kind}

The following useful relations have been found, which can be easily verified:

$$ \left(2z_N^\theta\right)^{N-1} J_N^2\left(z_N^\theta\right)sin{\theta\over N} = sin \theta. \eqno(5.1) $$

or
$$  J^2_N (z) = \,{1\over \left(2z_N^\theta\right)^{N-1} } \,{sin{\theta}\over sin(\theta / N)}    \eqno(5.1a) $$

It folows immediately. that zeros of $J_N(z)$ occur at $\theta = m \pi$, m being any intger. Roots of $J_N^2(z)$ are $cos(m \pi /N)$.

Obviously, the right side of these equalities equals zero at $\theta = \pi$, but $sin (\pi/N)$ is different from zero for any integer $N\geq 2$.
Therefore, the polynomial in $cos(\pi/N)$ in the left side of these equalities should be equal to zero. 
These relations provide the link between the general case considered at the beginning of section 4 and the particular 
case of the optimal kinematics with all scattering angles equal to $\theta/N$.

The known in mathematics Chebyshev polynomials of 2-d kind \cite{cubic,cheb-engl,cheb-r} 
\footnote{According to \cite{cheb-r}, the Chebyshev polynomials of 2-d kind have been considered first by his pupils A.Korkin and E.Zolotarev 
and were named in honor of their teacher. Therefore, it is correct to name these polynomials Chebyshev-Korkin-Zolotarev, 
or CKZ-polynomials.} are defined as
$$ U_n[cos\,\theta] = {sin (n+1)\theta \over sin \theta}. \eqno(5.2) $$
For any number $n$ these polynomials are defined as function of common variable $x = cos\,\theta$ which is confined in the interval
$ -1 \leq x \leq 1$. 
The recurrent relation
$$U_{n+1}(x) = 2x U_n(x) \,-\,U_{n-1}(x) \eqno(5.3) $$
with $x=cos\theta$, can be easily checked using this definition .
Indeed, $(5.2)$ can be written as
$$U_{n+1}(x)+ U_{n-1}(x) = 2x U_n(x),  \eqno(5.3a) $$
or
$$ sin (\alpha +\theta) + sin (\alpha -\theta) = 2 sin\alpha\, cos \theta \eqno(5.3b) $$
with $\alpha = (n+1) \theta$, which is well known trigonometrical relation.
Several examples are presented in the table.
Different equivalent general expressions for the CKZ-polynomials are presented in \cite{cheb-engl}:
$$ U_n(x) = \sum_{k=0}^{[n/2]}C_{n-k}^k\, (2x)^{n-2k} = $$
$$ = \sum_{k=0}^{[n/2]} C_{n+1}^{2k+1} (x^2-1)^k x^{n-2k} = \sum_{k=0}^{[n/2]}C_{2k-n-1}^k (2x)^{n-2k}, $$
where $n>0$,  $[n/2]$ is the integer part of $n/2$, $C_n^m = n!/[m!(n-m)!]$ is the number of combinations.
The first of these formulas coincides with the expression, presented in \cite{long}
up to some coefficient.

It follows from above definition that zeros (roots) of polynomials take place when $sin (n+1)\theta =0$, but  $sin \theta $ is different from
zero. So, there are $n$ roots at 
$$\theta = {\pi \over n+1}, \theta = {2\pi \over n+1}, ...\quad \theta = {n\pi \over n+1},   \eqno(5.4) $$
therefore we have the equations which define the values of $cos (k\pi/n)$ at arbitrary integer $n$ and $k$.

The ortonormality conditions for the CKZ polynomials have the form
$$ \int_{-1}^1 U_m(x) U_n(x) = \pi \delta_{nm} \eqno(5.5)$$
which can be eazily verified using the trigonometrical definition of these polynomials.
 
The relation between characteristic polynomials $J^2(cos(\theta/N))$ and CKZ-polynomials $U_N(cos(\theta/N))$ takes place
$$ \left(2z_N^\theta\right)^{N-1} J_N^2 = U_{N-1} (z_N^\theta) = {sin\,\theta \over sin (\theta/N)}.  \eqno (5.6)$$
with $z_N^\theta = cos (\theta/N)$.
It is sufficient to check this relation for some small values of $N$, e.g. $N=2$ and $3$ we have on the left side
$2 cos (\theta/2)$ and $4 cos^2 (\theta/3)$, and on the right side we have $sin \theta /sin(\theta/2) =2cos(\theta/2)$ and
$sin \theta/ sin(\theta/3) = 4cos^2\theta - 1$. For greater $N$ the equality will be valid 
because recurrency relations for the right and left sides are essentially the same.
This connection was written explicitly in \cite{chebkop}. 
                                     
\begin{center}
\begin{tabular}{|l|l|l|l|l|l|l|l|}                   
\hline
$N$& $J_N^2(z_N)$& $ U_{N-1} (x)$ \\
\hline
3 &$1 - 1/4x^2$  & $4x^2 -1 $     \\
\hline
4 &$ 1-1/2x^2 $&$ 8x^3 -4 x $   \\
\hline
5 &$ 1-3/4x^2 +1/16x^4$  & $16x^4 -12 x^2 +1 $ \\
\hline
6 &$ 1-1/x^2 +3/16x^4$  & $32x^5 -32 x^3 +6x  $ \\
\hline
7 &$ 1-5/4x^2 +3/8x^4 -1 / 64 x^6$  & $64x^6 -80 x^4 + 24 x^2  -1 $ \\
\hline
8 &$ 1-3/2x^2 +5/8x^4-1 /16 x^6$  & $128x^7 -192 x^5 +80 x^3 - 8 x $ \\
\hline
8 &$ 1-7/4x^2 +15/16x^4-5 /32 x^6 +1/256 x^8 $  & $256x^8 -448 x^6 + 240 x^4 - 40 x^2 $ + 1 \\
\hline
\end{tabular}
\end{center}

 Table. Characteristic polynomials $J_N^2 (x)$ presented in \cite{koma1,koma2} and Chebyshev polynomials of 2-d kind $U_{N-1}$ given in literature 
\cite{cubic,cheb-engl,cheb-r}.
$x=cos (\theta/N)$. The connection $ U_{N-1}(x) = (2x)^{N-1} J_N^2(x)$ can be easily verified.\\


Evidently, zeros of $J_N(z)$ occur at $\theta = m \pi$, $m$ being any integer, and
roots of $J_N^2(z)$ are $cos (m\pi/N)$.

It follows from here that $lim  J_N^2|_{\theta \to 0}  = N/ 2^{N-1} $.

Now, with the help of Eq. $(5.6)$, we can cross-check the relations obtained in previous section, using known trigonometrical identities.
Let us denote  $z=cos \phi$. Then, Eq. $(4.20)$ can be rewritten in trigonometrical form

$$ {sin N\phi \over sin\phi(2cos\phi)^{N-1}} = {sin (N-k)\phi\over sin\phi (2cos\phi)^{N-k-1}} {sin(k+1)\phi\over sin\phi (2cos\phi)^k } -
{sin (N-k-1)\phi\over sin\phi (2cos\phi)^{N-k}} {sin k\phi\over sin\phi (2cos\phi)^{k-1} } .   \eqno (5.7) $$
After removal of some common factors in left and right sides, we come to the equality to be checked

$$  sin N\phi \,sin \phi = sin (N-k)\phi \;sin \,(k+1) \phi  - sin (N-k-1)\phi\;sin\, k\phi, $$
which can be checked using known relations $sin(N-k)\phi = sin\,N\phi \;cos\,k\phi -cos\,N\phi \;sin\,k\phi, $ etc.

We can check other relations in similar way. From 
$$J_{2m}^2 = {sin(2m\phi) \over sin\phi \;  (2cos\phi)^{2m-1}}; \; J_{m+1}^2 = {sin(m+1)\phi \over sin\phi\;  (2cos\phi)^m};\;
J_{m-1}^2 = {sin((m-1)\phi \over sin\phi\; (2cos\phi)^{m-2}} \eqno (5.8) $$
and  takes the form
$$ {sin(2m\phi) \over sin\phi\; (2cos\phi)^{2m-1}} = {sin(m\phi) \over sin\phi \;(2cos\phi)^{m-1}}
 \left[{sin(m+1)\phi \over sin\phi\; (2cos\phi)^{m}} - {sin(m-1)\phi \over sin\phi\; (2cos\phi)^m}\right], \eqno(5.9) $$
and after cancellation of common factors we come to

$$ sin(2m\phi) = {sin(m\phi)\over sin \phi} \left[sin(m+1)\phi - sin(m-1)\phi \right], \eqno(5.10) $$
which can be easily verified using well known trigonometrical relations.

This relation  can be cross-checked in similar way. 

\section{The backward focusing effect (Buddha's light of cumulative particles)}
This is the sharp enhancement of the production cross section at the strictly 
backward direction, $\theta = \pi$. 
This effect has been noted first experimentally in Dubna (incident protons, final particles protons and deuterons) \cite{foc-1}
and somewhat later by Leksin's group (incident protons of 7.5 Gev/c, emitted protons
of 0.5 Gev/c) \cite{foc-2}.
This striking effect was not well studied previously, both experimentally
and theoretically.
In the papers \cite{kop2,long} where the small phase space method has been developed, 
it was noted that this effect can appear due to multiple interaction processes. However, the
consideration of this effect was not detailed enough, and estimates have not been made
\footnote{\small At 80-th I have discussed mechanisms of cumulative production 
with professor Ya.A.Smorodinsky
who noted its analogy with known optical phenomenon - glory, or "Buddha's light". The
glory effect has been mentioned by Leksin and collaborators \cite{glo-1}, however,
it was not clear to authors of \cite{glo-1}, can it be related to cumulative production,
or not. In the case of the optical (atmospheric) glory phenomenon the light scatterings 
take place within drolets of
water, or another liquid. A variant of the atmospheric glory theory can be found in \cite{khare}. 
However, the optical glory is still not fully understood, existing explanation 
is still incomplete, see, e.g. http://www.atoptics.co.uk/droplets/glofeat.htm. In nuclear physics
the glory-like phenomenon due to Coulomb interaction has been studied in \cite{maiorova} for the case of low energy
antiprotons (energy up to few $KeV$) interacting with heavy nuclei.}.

Mathematically the focusing effect comes from the consideration of the phase space of 
the whole process in the method of the small phase space adequate in
this case. It takes place for any multiple interaction process, regardless the particular
kind of particles or resonances in the intermediate states.
When the angle of cumulative particle emission is large, but different from $\theta = \pi$,
there is a prefered plane for the whole process, as it was explained in previous section.
The deviations of real angles of particles in intermediate states, including all azimuthal angles,
 from the optimal, or basic kinematics with $\phi_k=0,\, \theta_k=\theta/N$ are small.
When the final angle $\theta = \pi$, then integration over one of azimuthal angles takes place
for the whole interval $[0, 2\pi]$, which leads to the rapid increase of the resulting cross section.

General case is of special interest.
The particular values of the polar angles of the momenta in the rescattering process - which correspond to the maximal momentum 
of the final emitted particle - depend on the masses of the rescattered objects. It can be proved, however, that the backward focusing
effect takes place for any values of the polar angles.
For arbitrary polar angles $\theta_k$ 
$$z_k= (\vec n_k \vec n_{k-1}) \simeq cos (\theta_k - \theta_{k- 1}) (1-\vartheta_k^2/2) -sin ((\theta_k - \theta_{k- 1})) \vartheta_k  + $$ 
$$  + sin (\theta_k) sin \theta_{k-1} (\varphi_k -\varphi_{k-1})^2/2. \eqno (6.1)$$

After substitution $sin \theta_k\varphi_k \to \varphi_k$ we obtain

$$z_k= (\vec n_k \vec n_{k-1}) \simeq cos (\theta_k - \theta_{k- 1}) (1-\vartheta_k^2/2) -sin ((\theta_k - \theta_{k- 1})) \vartheta_k 
+ {s_{k-1}\over 2s_k}\varphi_k^2 + $$
$$ + {s_{k}\over 2s_{k-1}}\varphi_{k-1}^2-\varphi_{k-1})\varphi_k, \eqno (6.2)  $$
where we introduced shorter notations $s_k = sin \theta_k$.

The quadratic form depending on the azimuthal angles $\varphi_k$ for the $N$-fold process is
$$Q_N(\varphi_,\varphi_l) = {s_2\over s_1}\varphi_1^2 + {s_1+s_3\over s_2}\varphi_2^2 +{s_2+s_4\over s_3}\varphi_3^2 +....
+ {s_{N-2}+s_N\over s_{N-1}}\varphi_{N-1}^2 -$$
$$-2 \varphi_1\varphi_2 -2\varphi_2\varphi_3 - ... - 2\varphi_{N-2}\varphi_{N-1}, $$
with $s_N= sin \theta$.

E.g., for $N=5$ we have the matrix

$$||a||_{N=5}(\theta_1,\theta_2,\theta_3,\theta_4)=
\left[\begin {array}{cccc} s_{\theta_2}/s_{\theta_1} & -1 & 0 & 0\\ 
-1 & (s_{\theta_1}+s_{\theta_3})/s_{\theta_2} & -1  & 0   \\
 0 & -1  & (s_{\theta_2}+s_{\theta_4})/s_{\theta_3}& -1\\
0  &  0 & -1 & (s_{\theta_3}+s_{\theta})/s_{\theta_4}
   \end {array}\right], \eqno (6.3  )$$

Determinant of this matrix can be easily calculated:

$$ Det \left(||a||_{N=5}\right) = J^2_5 = {s_\theta\over s_{\theta_1}} \eqno(6.4) $$

It can be shown further by induction that at arbitrary $N$
$$ Det \left(||a||_N\right) = J^2_N = {s_\theta\over s_{\theta_1}} \eqno(6.5) $$

It follows from the expression $(3.2)$ for the matrix $||a||$
$$Det ||a||_{N+1} (\theta) = {s_{N-1}+s_\theta\over s_N} Det(||a||_N(s_N)- Det \left(||a||_{N-1}\right)(s_{N-1}) \eqno (6.6) $$
Since $Det (||a||_N(s_N) = s_N/ s_1 $ and $Det (||a||_N(s_{N-1}) = s_{N-1}/ s_1 $,
we obtain easily
$$Det ||a||_{N+1} (\theta) = {s_{N-1}+s_\theta\over s_N}{s_N\over s_1} - {s_{N-1}\over s_1} = {s_\theta \over s_1} \eqno(6.7) $$

\section{Cross section at singular point $\theta = \pi$}

For particles emitted strictly backwards the phase space has different form, instead of
$J_N(\theta/N)$ enters $J_{N-1}(\theta/N)$ which is different from zero at $\theta =\pi$, and we have instead of Eq. (3.6)
$$ I_N(\varphi, \vartheta) = \int \delta\biggl[\Delta^{ext} - z_N\bigg(\sum_{k=1}^{k=N} \varphi_k^2 -\varphi_k\varphi_{k-1}/z_N
+\vartheta_k^2/2\biggr)\biggr] \left[\prod_{l=1}^{N-2} d\varphi_ld\vartheta_l\right] 2\pi d\vartheta_{N-1} = $$
$$= \frac{\left(\Delta^{ext}\right)^{N-5/2} (2\sqrt 2 \pi)^{N-1}}{J_{N-1}(z_N) \sqrt N (2N-5)!! z_N^{N-3/2}}, \eqno (7.1) $$

This follows from the expression for the quadratic form in azimuthal angles, where $J_N(\pi/N) =0$, and integration over 
$\phi_{N-1}$ takes place over the whole interval $[0, 2\pi]$.

To illustrate the axial focusing which takes place near $\theta = \pi$ the ratio is useful of 
the phase spaces at $\theta =\pi$ and near the backward direction. The ratio of the observed cross sections in the interval
of several degrees slightly depends on the elementary cross sections and is defined mainly by this ratio.
It is
$$ R_N = {\Phi(z)\over \Phi (\theta =\pi)} = \sqrt{\Delta^{ext}\over z_N}  {(2n-5)!! \sqrt N \over 2^{N-1} (N-2)!!}{J_{N-1}(z_N)\over J_N(z)} 
\eqno (7.2)  $$
Near $\theta = \pi$ we use that 
$$J_N(z) \simeq \sqrt{[J_N^2]'(z_N) sin{\pi\over N}{\pi - \theta\over N} } \eqno (7.3) $$
and thus we get
$$ R_N = C_N \sqrt{{\Delta_N^{ext}\over \pi - \theta}} \eqno (7.4) $$
with
$$C_N = {J_{N-1}(z_N)\over [(J_N^2)'(z_N) sin(\pi/N)]^{1/2}} {\sqrt N (2N-5)!! \over \sqrt{z_N} (N-2)!! 2^{N-1}} \eqno(7.5)  $$

The phase space for the case of nucleons
$$ \Phi_N(\theta = \pi) = {1+ \zeta_1^2 \over k(m+\omega_{N-1})\zeta_N(1-\zeta_1^2)} \left( {\sqrt 2 \pi\over \zeta_{N-1}}\right)^{N-1}
\frac {[\zeta_N - k/(\omega +m)]^{N-1}}{I_{N-1}(z) \sqrt N (6N-3)!!} \eqno (7.6) $$

There are other data where the glory like effect is clearly seen. There are also data where the cross section
enhancement near the final angle $\theta = \pi$ is not observed, but in such experiments  the deviation
of the final angle from 180 deg. usually is large, as e.g. in \cite{foc-2} .

\section{Conclusions}
The origin of cumulative particles is not quite clear yet. The well known statement could characterize situation with these studies:
"never had so many people in such a long period of time done so little success..."

The glory-like backward focusing effect, noted first experimentally at JINR \cite{foc-1}, somewhat later observed at ITEP \cite{foc-2,foc-4}, and studied 
in more details in \cite{glo-1,glo-2}, was considered as a puzzle.
We have shown that just consideration of the MIP contributions, very unpopular (absolutely out of fashion) during many years, allows, in principle,
to explain this effect.
Relatively simple physics argumentation, based on the small phase space method for 
description of the MIP probability in so called "kinematically forbidden regions" (cumulative 
particles production), leads to appearance of the characteristic polynomials in polar angles variables.

Some correct expressions for the probability of the MIP were obtained about 40 years ago \cite{kop2}, characteristic polynomials 
$J_N^2$ have been obtained
here for the multiplicity $N\leq 5$. General expression for $J_N^2[cos (\theta/N)]$ was presented in \cite{long} more than 30 years ago.
This is somewhat painstaking work, which did not attract much attention of theorists, working in the field of high energy nuclear physics.
And only recently connection between $J_N^2$ and known in mathematics Chebyshev polynomials of 2-d kind became clear \cite{chebkop}.
These polynomials, which should be called also the Chebyshev-Korkin-Zolotarev (CKZ) polynomials, are
widely used in approximation theory and in other fields \footnote{Chebyshev polynomials are heavily used in numerical solutions. 
One of well known applications is in electrical filters. 
The Chebyshev filter is arguably one of the most common filters used in any electrical circuit that demands a filter. 
There is a 90\% probability that the mobile device in our pocket uses a Chebyshev filter in either the audio or RF filter section. 
The benefit of the Chebyshev polynomial in a filter design is that it allows a much faster rolloff of the filter "skirts" 
compared to other filter polynomials (e.g. a Butterworth) by allowing some passband response ripple.
Chebyshev polynomials can be applied to the optimal control of time-varying linear systems.}.

The cross section enhancement near the backward direction
can be explained as a result of multistep processes contributions just because of remarkable properties
of the characteristic polynomials in angular variables, which enter the denominator of the expression for the probability of
the process ($d\sigma \sim 1/J_N[sin(\theta/N)]$). There is huge amount of different kinds of MIP, elastic and inelastic,
kinematical boundary some of them is near to the momentum of observed cumulative particle, and just such MIP will provide
a glory-like enhancement of the cross section. An actual problem is to observe this effect for different kinds of particles
--- cumulative kaons, hyperons, etc, not only pions and nucleons, and for different projectiles.

It is a pleasure for me to thank Stepan Shimansky, Egle Tomasi-Gustafsson for discussions of some questions. I appreciate also
help of Galina Matushko in preparation of some previous publications \cite{koma1, koma2}.

\section{Appendix. Useful mathematical relations and formulas}

We prove here the relation   (see Eq. (3.3) )
$$ {1\over \omega_N'}\delta(m +\omega_{N-1} -\omega - \omega_N') = {1\over k k_{N-1}} 
\delta \left({m\over k} - \sum_{n=1}^{n=N} (1-cos\,\theta_n) \right), \eqno(A.1) $$
$\omega = \omega_N$ is the energy of the final (cumulative) particle. 

We recall that for the case of the light particle $\omega = k =|\vec k|$, and $k_N = k$.
This relation converts the energy conservation in the last ($N$-th) interaction act into the restriction
on the angular deviations from the optimal kinematics of the MIP (the constraint for the quadratic form in angular deviations),
which is crucially important in further treatment.
The energy-momentum conservation takes place in each of interaction acts.

The recoil energy in the last interaction act equals
$$\omega_N' = \sqrt{m^2 +(\vec k_{N-1} -\vec k)^2 } =\sqrt{m^2+k_{N-1}^2 +k^2 -2 k k_{N-1} z_N } \eqno(A.2)$$

and we can find the value $z_N^0$ which satisfies the above $\delta$-function, i.e. the equality
$$ m+\omega_{N-1} = \omega +\omega_N' \eqno(A.3) $$
or
$$\omega_N'^2 = m^2+k_{N-1}^2 +k^2 -2 k k_{N-1} z_N = (m + \omega_{N-1} - \omega )^2.\eqno(A.4) $$
Taking into account that $\omega_{N-1} = k_{N-1}$, $\omega = k$,
we obtain easily
$$ 1-z_N^0 = { m(k_{N-1} - k)\over k k_{N-1}}.\eqno(A.5) $$

When $z_N$ is near $z_N^0$, we can write
$$ \delta [f(z_N)] = \delta [f_{z_N}' (z_N = z_N^0)] = {\omega_N'\over k k_{N-1}} \delta(z_N-z_N^0), \eqno (A.6) $$
where function $f = m+\omega_{N-1} - \omega - \omega_N'$, and 
$$f_{z_N}' (z_N = z_N^0) = {k k_{N-1} \over \omega_N'}, \eqno (A.7)$$
and we come to 
$$z_N - z_N^0 = {m\over k} - {m\over k_{N-1}} + z_N -1. $$
Taking into account that 
$$ {m\over k_{N-1}} = \sum_{l=1}^{l=N-1} (1 - z_l) $$
we come to eq. $(A.1)$.
The elastic rescattering in the first interaction act was assumed here. Not essential change should be done for the 
inelastic interaction.

Here we present for the readers convenience some formulas and relations which have been used in present paper (sections 3 and 4).
The nuclear glory phenomenon is an example when solving the physics problem leads to mathematical consequences of interest.

In sections 3 and 4 the following integrals of the $\delta$-functions have been used:
$$I_n(\Delta)=\int \delta(\Delta - x_1^2-... - x_n^2) dx_1... dx_n = \pi {(2\pi)^{(n-2)/2}\over (n-2)!!} \Delta^{(n-2)/2} \eqno(A.8)$$
for integer even $n$.

$$I_n(\Delta)_n= \int \delta(\Delta - x_1^2-... - x_n^2) dx_1... dx_n =  {(2\pi)^{(n-1)/2}\over (n-2)!!} \Delta^{(n-2)/2} \eqno(A.9)$$
for integer odd $n$.
Relations
$$ \int_0^\pi sin^{2m}\theta\;d\theta\,=\,\pi{(2m-1)!!\over (2m)!!}; 
\qquad \int_0^\pi sin^{2m-1}\theta\;d\theta\,=\, 2{(2m-2)!!\over (2m-1)!!}, \eqno(A.10) $$
$m$ --- integer, allow to check $(6.1)$ and $(6.2)$ easily.

The equality takes place
$$\int \delta(\Delta - x_1^2-... - x_n^2)\delta(x_1+x_2+... +x_n) dx_1... dx_{n-1} dx_n= 
 {1\over \sqrt n}  I_{n-1}(\Delta)  \eqno(A.11)$$
More generally, for any quatratic form in variables $x_k, \; k=1, ... n$ after diagonalization we obtain  
$$ \int \delta(\Delta - a_{kl}x_kx_l)dx_1\, ... \, dx_n= 
\int \delta(\Delta - x_1'^2-... - x_n'^2) {dx_1'...dx_n' \over \sqrt{det ||a||}} =
{1 \over \sqrt{det ||a||}}I_n(\Delta). \eqno(A.12) $$

The polynomials $J_N^2$ and equations for $z_N^\pi=cos(\pi/N)$ can be obtained in more conventional way.
There is an obvious equality
$$ [exp (i\pi/N)]^N = exp(i\pi) = -1 \eqno(A.13)$$
It can be written in the form                                            
$$[cos(\pi/N) +i sin (\pi/N)]^N = -1, \eqno(A.14)$$
or separately for the real and imaginary parts
$$Re \left\{ [cos(\pi/N) +i sin (\pi/N)]^N\right\} = -1,\quad Im \left\{ [cos(\pi/N) +i sin (\pi/N)]^N\right\} = 0. \eqno(A.15)$$
The polynomials in $z_N^\pi=cos(\pi/N)$ which are obtained in the left side of $(6.13)$ coincide with polynomials obtained in section 4.
However, some further efforts are necessary to get recurrent relations $(6.9),\,(6.10)$.

\vglue 0.2cm

\end{document}